\documentclass[pdflatex,sn-mathphys-num]{sn-jnl}

\usepackage{graphicx}%
\usepackage{multirow}%
\usepackage{amsmath,amssymb,amsfonts}%
\usepackage{amsthm}%
\usepackage{mathrsfs}%
\usepackage[title]{appendix}%
\usepackage{xcolor}%
\usepackage{textcomp}%
\usepackage{manyfoot}%
\usepackage{booktabs}%
\usepackage{algorithm}%
\usepackage{algorithmicx}%
\usepackage{algpseudocode}%
\usepackage{listings}%
\usepackage{dcolumn}
\usepackage{bm}
\usepackage{multirow}
\usepackage{svg}
\usepackage{tabularx}

\newcommand{\karim}[1]{{\color{black}#1}}
\newcommand{\water}{$\mathrm{H_2O}$}

\begin{document}

\title{Overfitting by design: neural network density functionals for water}

\author*[1]{\fnm{Karim K.} \sur{Alaa El-Din}}\email{karim.alaael-din@physics.ox.ac.uk}

\author[1]{\fnm{Antonius} \sur{v. Strachwitz}}

\author[1]{\fnm{Ana} \sur{Coutinho Dutra}}

\author[1, 2]{\fnm{Sam M.} \sur{Vinko}}

\affil*[1]{\orgdiv{Department of Physics}, \orgname{University of Oxford}, \orgaddress{\street{Clarendon Laboratory, Parks Road}, \city{Oxford} \postcode{OX1 3PU}, \country{UK}}}

\affil[2]{\orgdiv{Central Laser Facility}, \orgname{STFC Rutherford Appleton Laboratory}, \orgaddress{\city{Didcot} \postcode{OX11 0QX}, \country{UK}}}

\date{\today}

\abstract{
In density functional theory, simpler exchange-correlation (XC) approximations such as the local density approximation (LDA) are favored for computational speed but rely on limited information, leading to a trade-off between accuracy and generality. Machine-learned XC approximations have seen a lot of interest to address this problem. Here, we train a neural network LDA using a differentiable Kohn–Sham solver, imparting system-specific expertise for water and sacrificing generality for accuracy. Our model achieves \karim{1 kcal / mol errors} on gold standard coupled cluster ionization and atomization energies, and improves predictions of spectral lines, electron density distribution, and equilibrium geometry from as few as eight configurations used for training. We proceed to perform transfer learning and obtain results comparable to higher-rung PBE and B3LYP functionals on the WATER27 subset of the GMTKN55 database, even when only a single two-molecule binding energy is used in the transfer process. This result opens the door for specialist functionals to be trained on different systems from little data, enhancing predictions while maintaining low training costs. Our approach of training a modified XC density functional approximation (DFA) furthermore allows for a highly interpretable result, as the neural network directly corresponds to a correction of the XC energy per electron.
}

\maketitle

\section{Introduction}

In the sixty years since its original inception, density functional theory (DFT) has seen an astonishing range of applications~\cite{Vinko2014DensityPlasmas, Orio2009DensityTheory, Neugebauer2013DensityScience}, driven by the high accuracy at relatively low cost inherent to this framework~\cite{Burke2012PerspectiveTheory}. From modeling planetary interiors~\cite{Oganov2005AbMaterials} to hydrogen storage~\cite{Suleyman2009DFTStorage} and chemical receptor binding studies~\cite{Kangas1998OptimizingProperties}, few tools have proven as influential as DFT in physics, material science and beyond. Thus, there is a sustained interest in possible improvements in accuracy or computational speed~\cite{Sim2022ImprovingTheory, Trushin2025ImprovingEnergies}, which could open the door to novel applications, and accelerate existing use cases of DFT.

Much of this interest has focused on the exchange-correlation (XC) functional, which encodes the many-body effects of an interacting Coulomb system. In principle, a universal XC term exists  but it cannot be computed efficiently~\cite{Schuch2009ComputationalFunctionaltheory} and in practice, a range of density functional approximations (DFAs) are used instead~\cite{Goerigk2017AInteractions}.
These range from purely local approximations~\cite{Perdew1992AccurateEnergy} to semi-local~\cite{Perdew1996GeneralizedSystem} and hybrid approaches~\cite{Becke1993ATheories, Becke1993DensityfunctionalExchange} which incorporate additional information and correspondingly tend to be both more accurate and expensive. Collectively, these approximations are commonly referred to as rungs on Jacob's ladder~\cite{Perdew2001JacobsEnergy}, with each rung containing different DFAs utilizing the same amount of input information. In total, there are over 700 DFAs reported in the library of XC functionals (LibXC)~\cite{Lehtola2018RecentTheory}, with each corresponding to a different parametrization of the XC term. Such parameterizations may be guided either by non-empirical derivations, or empirical fitting processes.

Recently, machine-learned DFAs have seen a surge of interest~\cite{Fiedler2022DeepChemistry, Pederson2022MachineTheory,  Kulik2025AreChemist, Akashi2025CanDFT}, but have failed to see broad adoption  despite the promising results they deliver~\cite{Nagai2020CompletingMolecules, Kasim2021LearningTheory, Li2021Kohn-ShamPhysics, Kirkpatrick2021PushingProblem, Khan2025AdaptingLearning}. In many cases, these approaches have been employed to find a better approximation to the universal XC functional at a given rung of Jacob's ladder. In other words, the goal is to advance the Pareto front across all of chemical space without increasing the complexity of the DFA parameterization. However, it is unclear whether such a general advancement is possible at each rung, especially for the non-hybrid semi-local rungs constituted by local density approximations (LDAs), generalized gradient approximations (GGAs), and Meta-GGAs. The fact that no model has replaced the Perdew-Burke-Enzerhofer (PBE) GGA~\cite{Perdew1996GeneralizedSimple} and Perdew-Wang (PW) LDA~\cite{Perdew1992AccurateEnergy} on their respective rungs, despite concerted efforts to find more effective models across the past three decades, would suggest that any improvement across all of chemical space, if possible, would be small. Nevertheless, the popularity of these lower-rung models clearly highlights the importance of keeping computations fast and scaling favourably, particularly for modelling larger systems.

We show that improvements in accuracy can be made in this regime from very little data by sacrificing generalizability across chemical space, if an appropriate approach is chosen. To this end, we leverage a differentiable Kohn-Sham solver and the surrogate training embedded in physics (STEP) paradigm~\cite{AlaaEl-Din2024STEP:Learning}. This extends recent work by von Strachwitz \textit{et al.}~\cite{vonStrachwitz2026Data-efficientDFT}, demonstrating the usage of STEP-DFT to reconstruct known functionals from little data. We develop a specialist density functional approximation to characterize both single- and multi-molecule interactions in water, a system both ubiquitous and notoriously challenging to model due to the interplay of strong hydrogen bonds and weak van der Waals interaction. First, we construct a neural LDA trained on single-molecule properties of water obtained from only eight coupled cluster with singlet, doublet, and pertubative triplet (CCSD(T)) calculations, and demonstrate that the training properties can be learned to \karim{within 1 kcal / mol} across a wide range of configurations, while simultaneously optimizing the electronic density of the system. \karim{We selected 1 kcal / mol as the desirable error level, as this constitutes chemical accuracy when comparisons are made against gold-standard calculations}. In addition, we find that our trained DFA reproduces the correct density with much higher accuracy than other, higher-rung DFAs, indicating that both density-driven and functional driven errors are addressed. \karim{Note that these gains are due to the restrictions in chemical space to which the model is applicable, and thus constitute a departure from the uniform electron gas limit which typically serves as the justification for LDA models. Our work constitutes a purely heuristic set of models.} We then proceed to consider multi-molecule systems from the WATER27 subset of the GMTKN55 database, and apply transfer learning from the single-molecule model in combination with a single scalar CCSD(T) two body energy to realize large gains in performance across the entire dataset containing clusters of up to 20 molecules. We demonstrate that these gains persist in the complete basis set (CBS) limit, and that pre-training on single-molecule is required to realize them. We furthermore show that the single-molecule model shows no significant decrease in performance on the WATER27 dataset compared to the baseline PW-LDA, indicating that even outside their narrow training domain, specialist DFAs may remain suitable.

Although a growing body of similar efforts exists, there are key differences between our approach and other work published to date. First, our use of a small, simple feed-forward neural network as an additive correction to the XC term with the same input parametrization greatly reduces the complexity of the DFA compared to Kirkpatrick~\textit{et al.}~\cite{Kirkpatrick2021PushingProblem}, who rely on inclusion of Hartree-Fock energy densities and use a much larger dataset to construct a generalist DFA. Dick~\textit{et al.}~\cite{Dick2020MachineDensity} similarly rely on a careful procedure of learned projectors and symmetrizers, while restricting themselves to single molecule systems. Bogojeski ~\textit{et al.}~\cite{Bogojeski2020QuantumLearning} also restrict themselves to single molecule systems, and only consider total energies of the system for evaluation, although they similarly focus on smaller training sets. On the other hand, Fritz~\textit{et al.}~\cite{Fritz2016OptimizationWater} start from a van der Waals DFA which includes non-local interactions and use Bayesian optimization to fit a large dataset of dimer and trimer interaction energies in water. Furthermore, Kasim~\&~Vinko~\cite{Kasim2021LearningTheory} build a generalist model only investigating a few properties, while Li~\textit{et al.}~\cite{Li2021Kohn-ShamPhysics} restrict themselves to hydrogen in one dimension. Overall, the simplicity of our model and small dataset required for effective training of a specialist DFA distinguish this work from previous research. Finally, machine learned potentials provide a different avenue to enhanced predictions of molecular structure, particularly in a molecular dynamics context~\cite{Deringer2019MachineScience, Zuo2020PerformancePotentials}. However, they also require large datasets for training and are usually bound by the quality of their training data derived from DFT, which can be a limiting factor~\cite{Kuryla2025HowDatasets}. Thus, better DFAs are of interest also to the machine learning potential community, as they can improve training data for potentials.

\begin{figure}[h!]
    \centering
    \includegraphics[width=.35\linewidth, clip, trim={0, 0, 950, 0}]{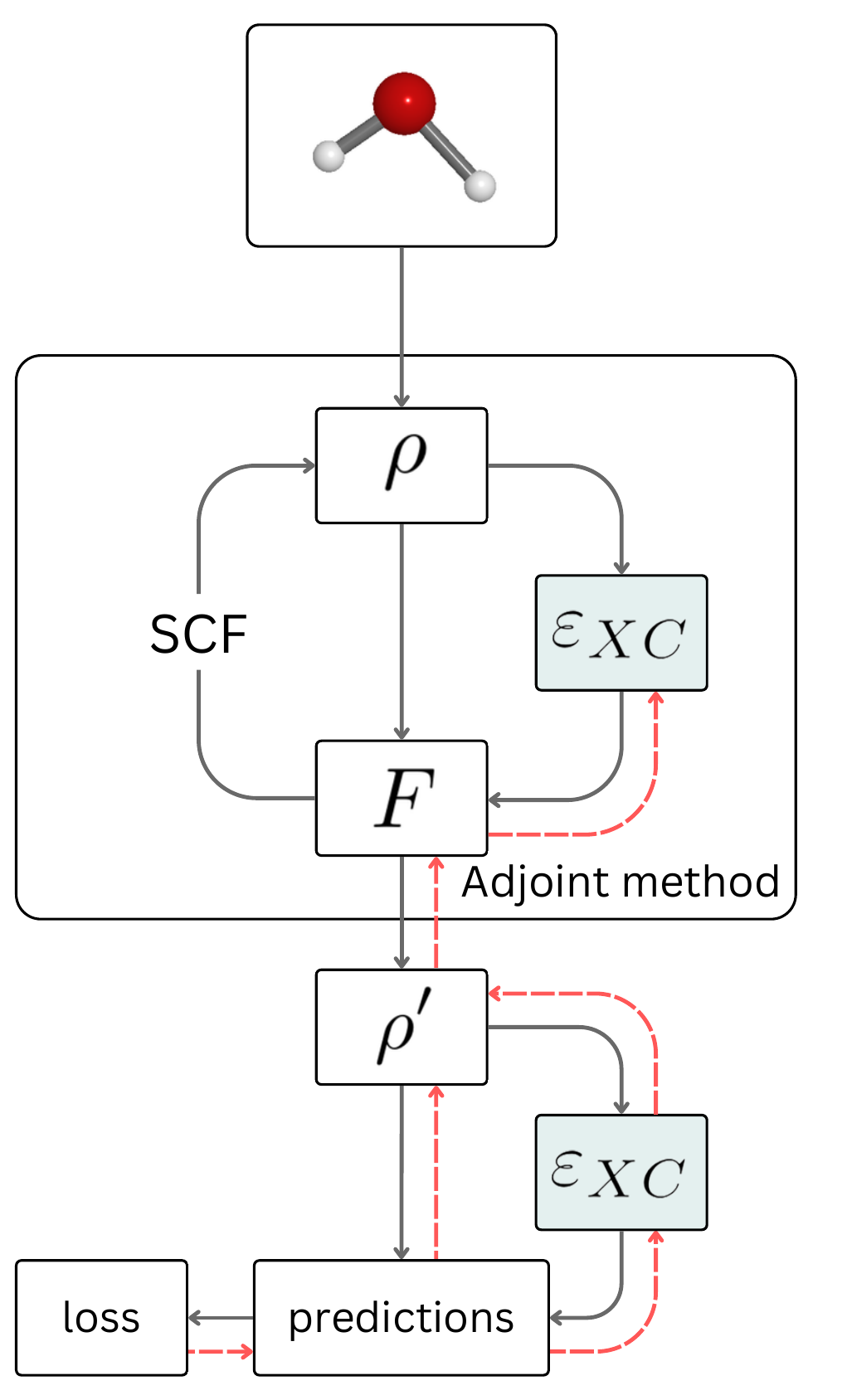}
    \caption{The algorithm used to train DFAs within a Kohn-Sham solver, as implemented by Kasim \textit{et al.}~\cite{Kasim2022DQC:Chemistry}. Black arrows indicate forward passes while red ones indicate backpropagation. $\epsilon_{XC}$ denotes the DFA which is optimized in this work.}
    \label{fig:scheme}
\end{figure}

\begin{figure*}[t!]
    \centering
    \includegraphics[width=\linewidth]{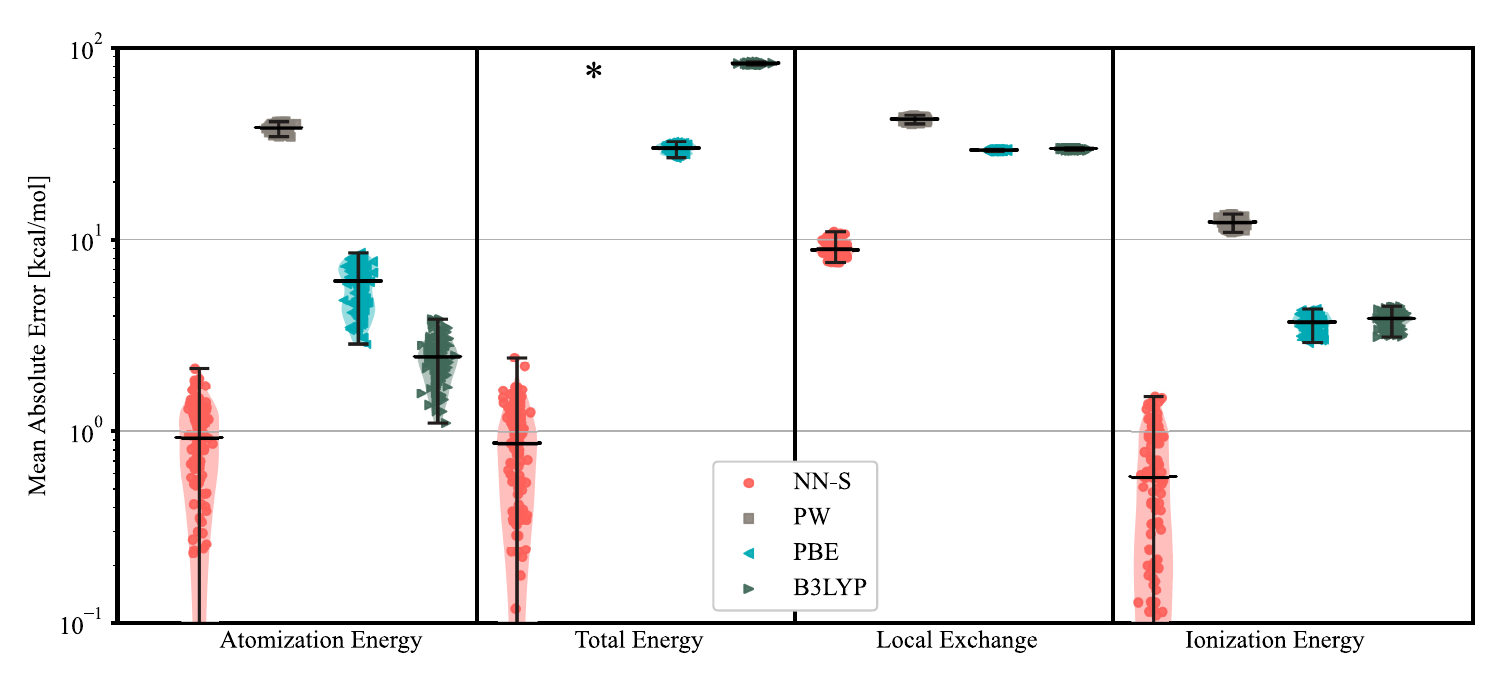}
    \caption{The mean absolute errors on the testing set of single molecule systems for different models. Note that the NN-LDA achieves \karim{errors below 1 kcal / mol} on the three energy properties (atomization, total, and ionization energies), while performing best on the local exchange loss (related to electron density distribution). The total energies for PW-LDA are omitted, as all errors are in excess of 100 kcal/mol. This is indicated with an $*$ in the figure.}
    \label{fig:inf_mono}
\end{figure*}

\section{Results}
\subsection{Training neural density functional approximations}
To perform training of the neural DFA, we utilize the differentiable quantum chemistry (DQC) code~\cite{Kasim2022DQC:Chemistry}. This implementation of Kohn-Sham DFT is fully differentiable and makes use of the adjoint method for computational efficacy, as illustrated in figure~\ref{fig:scheme}. We use a self-consistent field (SCF) solver to obtain converged estimates of the Fock matrix $F$ and the electron density $\rho$, from which our objectives are derived as detailed in section~\ref{meth:mol_calc}. Errors of these predictions may then be propagated backwards as losses, to obtain a corresponding error on the electron density and Fock matrix. While the exchange-correlation term contributes to these converged properties at every step of the SCF loop, we can leverage adjoint calculations as outlined in Kasim~\textit{et al.}~\cite{Kasim2020xi-torch:Library} to obtain the derivative of the converged Fock matrix with respect to the XC term. Overall, this greatly improves performance and allows for a dynamic and efficient Kohn-Sham solver design, without considerations for backpropagation through the loop.\\

We follow this design to train a neural field correction to an existing DFA at the relevant level of Jacob's ladder, the details of which can be found in section~\ref{meth:nn_dfa}. For the actual training cycle, we perform the KS calculation for each system at each epoch, backpropagating errors to update the DFA step by step. While expensive, this process is not prohibitive if training on small systems like those used here, and allows for direct optimization of the DFA with respect to objectives of interest.

\subsection{Application to single molecules of water}

We begin by training on the \water~geometries contained within the ANI1-ccx dataset~\cite{Smith2020TheMolecules}, details and a visual summary of which can be found in the supplemental material. In-house CCSD(T) calculations were  \karim{at the pc-2 basis level} and used to generate labels for atomization, ionization, and total energies, as well as a localized energy loss (LEL) calculated as

\begin{equation}\label{eq:lel}
    \mathrm{LEL} = \sum^N_{i=1}|\varepsilon_X(\rho_i) - \varepsilon_X(\tilde{\rho}_i)|V_i.
\end{equation}

\noindent{} Here, $\varepsilon_X$ is the well-known Slater exchange approximation, $i$ indexes the volume elements of the molecular grid, and $V_i$, $\rho_i$ correspond to the volume and electron density at the grid point respectively. Overall, \karim{this definition of LEL yields a non-unique} volume weighted mean absolute error of the electron density adjusted for the exchange energy contributed by a given density. \karim{The LEL is useful insofar as it ensures consistent scale between different error terms and avoids ad hoc weighting corrections. It is not to be confused with the LES established in the literature~\cite{Polak2025Real-spaceFunctionals}, which has a different definition, although both ensure locally correct behaviour in addition to molecular energetics.}

\begin{figure*}
    \centering
    \includegraphics[width=\linewidth, clip,trim={0, 100, 0, 100}]{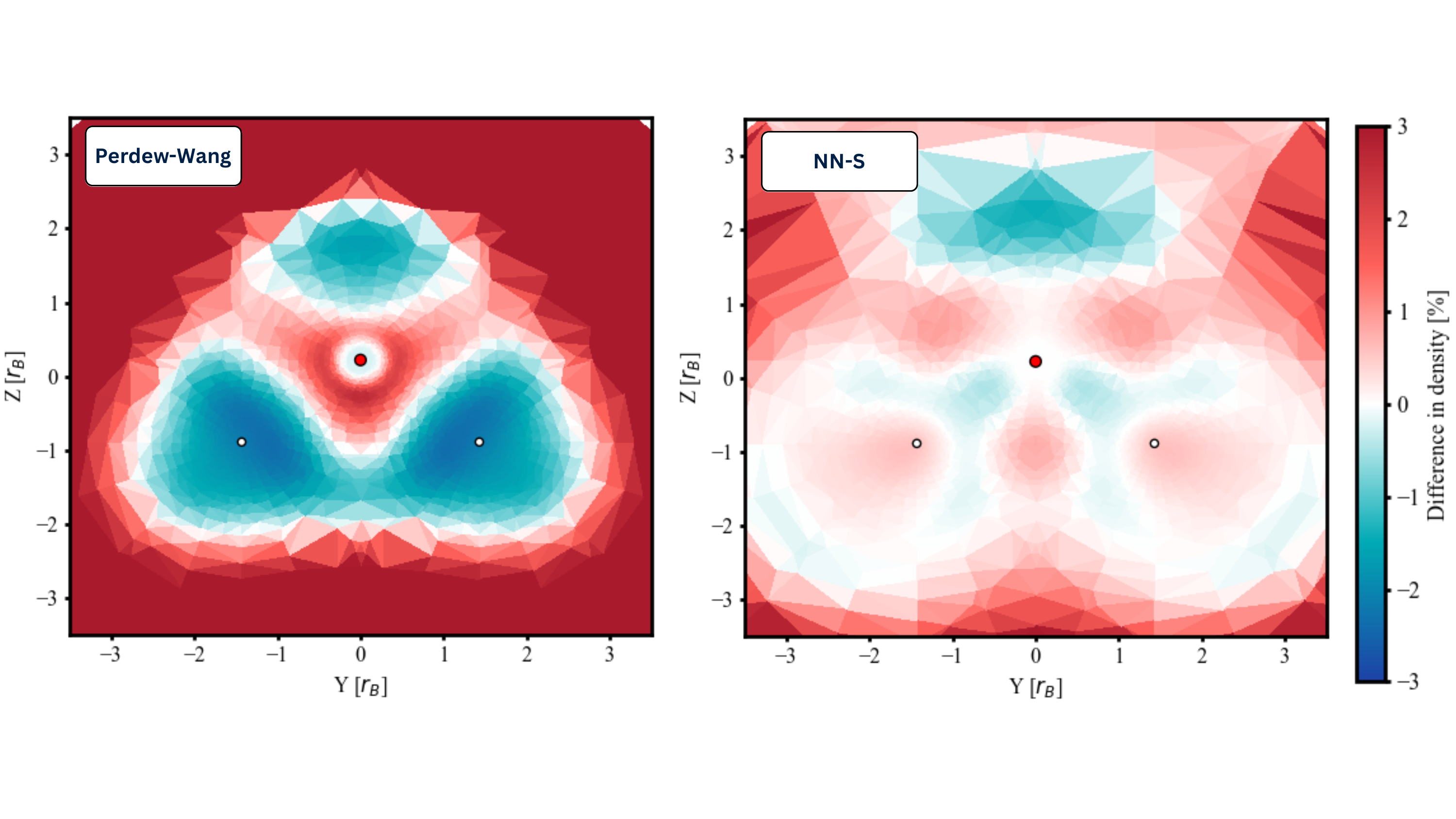}
    \caption{A side-by-side comparison of the electronic density difference w.r.t. CCSD(T) calculations on equilibrium \water. PW-LDA (left) exhibits significant (multiple \%) differences in density even in the key bonding regions. Our NN-S LDA (right) predicts densities much closer to the CCSD(T) baseline.}
    \label{fig:density}
\end{figure*}

Next, we train a LDA level correction to the PW approximation in the pc-2 basis~\cite{Jensen2001PolarizationPrinciples, Jensen2002PolarizationLimit}, performing delta learning to optimize a weighted combination of the aforementioned labels with 5 geometries for training and three further ones for validation. Details of data splitting can be found in the section 1 of the supplemental material. The resultant neural network DFA, which we call NN-S, exhibits \karim{errors below 1 kcal / mol} on all energetic properties across the remaining 112 geometries from the full dataset as seen in figure~\ref{fig:inf_mono}. This is notable, particularly in contrast to lower performance from the higher rung PBE and B3LYP functionals chosen for comparison. The improvement in electron densities with respect to PW-LDA is further highlighted in figure~\ref{fig:density}. \karim{While this set of comparison functionals is by no means exhaustive, it does include one of the best-performing functionals for water clusters in PBE. PBE outperforms not only B3LYP, but also dispersion corrected variants of both PBE and B3LYP as well as SCAN on the Water-27 dataset~\cite{Goerigk2017AInteractions}. Furthermore, other modern functionals such as PBE0, $\omega$-B97X-D3 or the double hybrid DSD-B3LYP have similar median absolute deviations to PBE on the Water-27 dataset, ranging from 2 to 3 kcal / mol~\cite{Goerigk2017AInteractions}. In addition, PBE and B3LYP are some of the most widely used functionals, and for applications where exhaustive studies on specialist functionals have not been done, these would also constitute the most likely starting points.}

While the performance on this dataset is interesting, it is not too surprising that we see improvements on properties present in the training set. Indeed, some of the more promising NN DFA approaches published in the past show overfitting to particular properties at least to some extent~\cite{Kasim2021LearningTheory}. However, we may furthermore consider properties outside the training set to assess performance of each DFA. When comparing to experimental results, we find that NN-S still outperforms base model PW-LDA across the board, and that it exceeds PBE performance across all properties but infrared intensities, as seen in the supplemental material. Thus, we can clearly see that NN-S is an improved DFA in the context of systems of single water molecules, and that we can realize an explicit gain by overfitting to a this specific chemical context. Crucially, overfitting is the correct term here, and although this is usually considered a negative, it can be beneficial in the STEP paradigm.

As long as the domain we overfit to is broad enough, we can view this simply as an improvement in accuracy, particularly as there is no noise attached to our training data. In our case, we overfit to the chemical context of molecular \water, and we have clear evidence that the resultant DFA is unphysical as shown in the supplemental material. However, the resultant DFA is optimized for \water, indicating that relevant systems are either insensitive to the unphysical characteristics with respect to the properties of interest, or that a distortion in contributed energies from these densities more accurately captures the nature of XC effects in water.

Overall, this overfitting clearly benefits calculations within the context of molecular \water, including improved bond breaking energies and ionization calculations. However, such a narrow scope is not particularly useful, as most interesting behaviour in water occurs at the intermolecular level. We thus turn to the WATER27 subset~\cite{Bryantsev2009EvaluationClusters} of the GMTKN55 database~\cite{Goerigk2017AInteractions}, to see how our original model performs on this data, and how gains on single-molecule systems may be transferred to clusters of multiple molecules.

\subsection{Transfer learning and application to water clusters}

Larger clusters of water molecules exhibit dispersion interactions that are notoriously difficult to capture with density functional theory\karim{. Thus, these systems typically warrant dispersion corrections and/or the use of higher rung DFAs, although neither guarantees improved performance.} Furthermore, they reach the scale of systems that become prohibitively expensive to model with CCSD(T) calculations. These two challenges make such systems particularly interesting for the application of neural specialist DFAs. However, as such a DFA needs to be retrained on every chemical context it is applied to, training on a large dataset of more extensive systems is undesirable. In an effort to maintain the comparatively low training cost of our single-molecule model, we thus continue with a low data approach and proceed to study how we can obtain accurate predictions on the WATER27 dataset with minimal data and only small systems in the training set.

The cheapest approach is to use the single-molecule NN-S model to extrapolate to multi-molecule systems. While there is no reason to assume that this model will perform well in this extrapolative regime, it is nevertheless instructive to see how it compares at least to PW-LDA when extrapolating to closely related systems. If such an extrapolation leads to significantly worse predictions than the base model, this would restrict specialist functionals only to systems well within their training domain.

Next, we consider the cheapest and easiest way of incorporating some information about multi-molecule effects into our neural DFA. While we could establish a large dataset of two-body and three-body interactions for training similar to~\cite{Fritz2016OptimizationWater}, this approach requires substantial datasets which exist for water, but not for other systems where specialist models are of interest. Instead, we use a transfer learning approach to add a small amount of multi-molecule information into our existing NN-S model. We retrain the model on the smallest $\mathrm{(H_2O)_2}$ entry of the WATER27 dataset until convergence, terminating the training before overfitting occurs. This happens very rapidly, within 20 epochs, due to the fact that we are training on a single scalar value. We dub the resultant model NN-T, and evaluate its performance on the remaining entries of the WATER27 dataset.

\begin{figure}
    \centering
    \includegraphics[width=.7\linewidth]{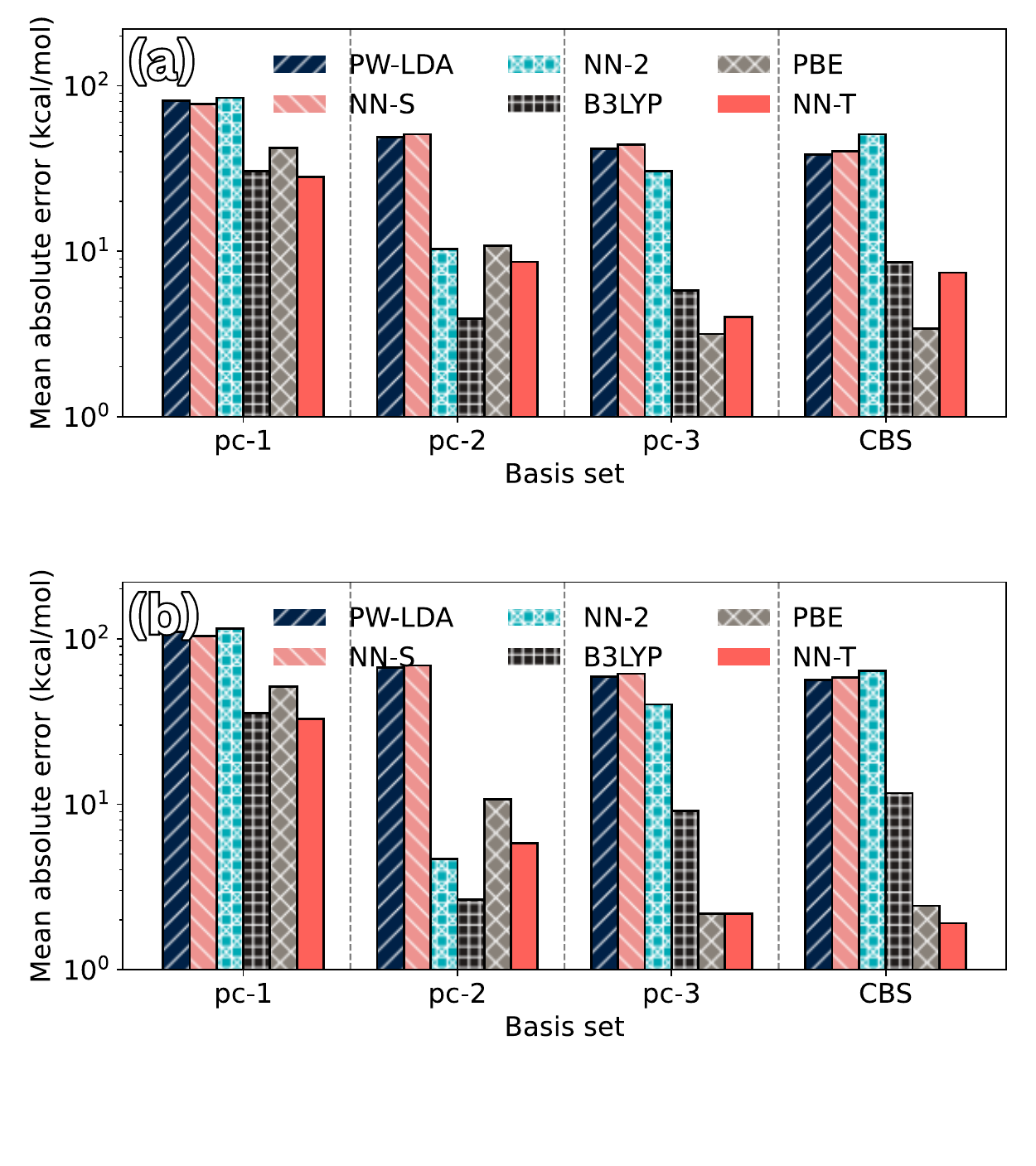}
    \caption{\karim{Mean absolute error of different models at different basis set accuracies on the WATER27 subset of GMTKN55 excluding the water dimer used for training of NN-2 and NN-T models. The pc-$\zeta$ are different polarization consistent basis sets, while CBS indicates the complete basis set limit. (a) Vertical bars indicate MAE on the entire remaining WATER27 subset. (b) Vertical bars indicate MAE only on the neutral subset of the remaining WATER27 data.}}
    \label{fig:inf-27}
\end{figure}

Finally, we compare these models against a third candidate model, which is similarly trained only on the $\mathrm{(H_2O)_2}$ entry of the WATER27 dataset, but beginning with a random initialization. This allows us to see whether the transfer learning is required, and whether we could even realize gains by learning from a single scalar.

All of these models are trained at pc-2 basis level, and the resultant models vary greatly in their exact functional form. A visual representation of this can be found in the supplemental material. In particular, NN-T retains a lot of features from NN-S, while NN-2 is completely distinct and has an unphysical sign across the entire density range encountered in water molecules. Notably, NN-T and NN-2 have converged to completely different local optima of the LDA space, indicating that transfer learning has had a pronounced effect.

Each model is now evaluated against reference CCSD(T) calculations from the GMTKN55 database. However, as the latter calculations are done in the complete basis set (CBS) limit, we calculate system energies using our three models in the pc-1, pc-2, and pc-3 bases and use the exponential square root Ansatz from Jensen~\cite{Jensen2005EstimatingCalculations} for extrapolation. We model the energy in the pc-$\gamma$ basis as 
\begin{equation}\label{eq:CBS}
    E_{\gamma} = E_{CBS} + A \cdot e^{-\beta \sqrt{\gamma -1}},
\end{equation}

with $\gamma - 1$ in the exponent, as the pc family is zero-indexed. $E_{CBS}$, $A$, and $\beta$ are parameters specific to each modeled system, which we solve for using the finite basis calculations for a given system. We may then take $E_{CBS}$ to be the CBS energy of the system.

The resulting mean absolute errors for each model and basis combination across the WATER27 dataset excluding the $\mathrm{(H_2O)_2}$ entry which was used for training are shown in figure~\ref{fig:inf-27}(a). A few trends emerge quite clearly. For one, the single-molecule NN-S model performs similar to the baseline PW, indicating that it has reasonable performance even in a slightly extrapolative regime. This increases our confidence in neural specialist DFAs as a more general approach, as such a model would be of little use if it failed completely when used even slightly outside the training domain. Secondly, we see that the transfer-learned NN-T model performs remarkably well, being the most accurate model at the pc-1 level, and the second most accurate at the pc-2 through CBS levels, despite having a substantially restricted input space compared to the hybrid B3LYP and the PBE-GGA DFAs. This is a notable result as it indicates that little data can suffice to unlock substantial performance gains in neural specialist DFAs if one does so carefully. The importance of careful training is further emphasized by the poor results from the two-body only NN-2 model, which performs much worse than its transfer-learned counterpart. We see reasonable improvements at the pc-2 level, where the NN-2 model outperforms PBE, but error levels are much higher for all other basis levels, yielding the worst performance of all DFAs under consideration at pc-1 and CBS levels. This behaviour is a telltale sign of the DFA compensating for the basis set error, which leads to underestimated energies as the complete basis set is approached, and indicates overfitting to a particular basis level. Similar but weaker effects can be seen for B3LYP, PBE and NN-T model, with B3LYP exhibiting the strongest effects. PBE errors rise only slightly from the pc-3 level to the CBS limit and interestingly NN-T performs best at pc-3 level despite being trained at pc-2 accuracy.

\begin{figure}
    \centering
    \includegraphics[width=\linewidth]{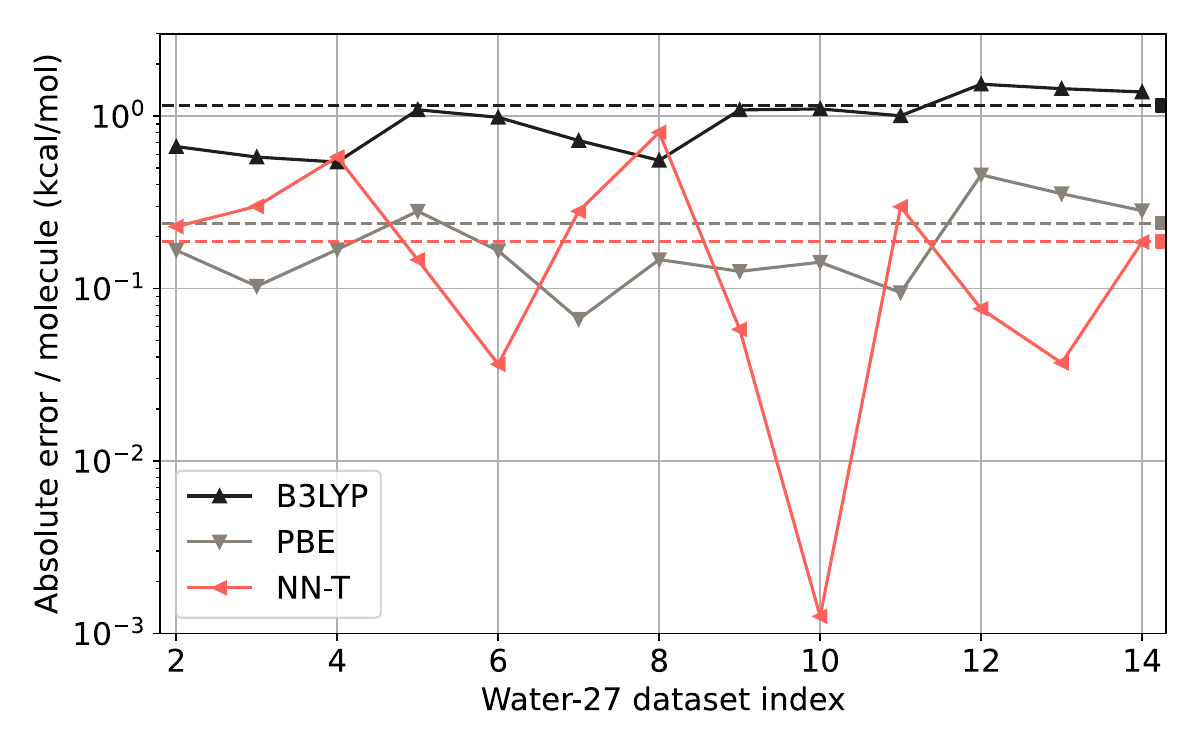}
    \caption{Absolute error per monomer of different models at the CBS limit on the neutral subset of the WATER27 dataset. Solid lines indicate the performance on each individual system, while the dashed lines indicate mean absolute error per molecule.}
    \label{fig:error_per_mol}
\end{figure}

These differing behaviours are explained in large part by the diverse makeup of the WATER27 dataset. The three main groups contained therein are clusters of neutral water molecules, protonated and deprotonated clusters. Evaluating performance on the neutral entries, which contains clusters of up to twenty molecules, in the CBS limit, we see that the transfer-learned NN-T is not only the best performing model on this dataset \karim{at pc-1, pc-3 and CBS levels as shown in figure~\ref{fig:inf-27}(b)}, but that the absolute error per monomer is also stable with system size as seen in figure~\ref{fig:error_per_mol}. This is a strong indicator of the model extrapolating well to even larger water clusters, in contrast to for example the DM21 model, which was found to have errors increasing with cluster size on a similar water dataset~\cite{Palos2022DensityFunctional}. \karim{Additionally, we note that the NN-T model is the only model monotonically improving on this subset with increasing basis completeness, and provides strong results across all basis set levels, making it suitable for use across basis sets.}

The main errors in the CBS limit for NN-T arise for protonated and deprotonated clusters, which is perhaps unsurprising for a model never exposed to any $\mathrm{H_3O+}$ or $\mathrm{OH-}$ ions. Interestingly, the structure of the error differs between protonated systems and deprotonated ones. In the former case, NN-T exhibits underprediction of energies at each basis level with errors increasing with basis set completeness, indicating overcompensation against basis set error. On the latter dataset however, we observe convergence towards the correct energies but with a high rate of convergence, which leads the exponential square root ansatz to underpredict CBS limit energies. This latter effect is likely an issue of this ansatz for systems containing $\mathrm{OH-}$, as all investigated DFAs exhibited this to some degree. Details of this effect can be found in the supplemental material. Finally, the last entry of the WATER27 dataset exhibits a combination of both effects, as it is the only one to feature both types of ions. Interestingly, the NN-T model here is the only one to approach the true energy with increasing basis accuracy, as all other DFAs investigated significantly underestimate the resulting energy. Ultimately, for this system NN-T exhibits a deprotonated signature while all others exhibit the underprediction characteristic of protonated systems. This is a curious result warranting further analysis of the CBS schemes chosen to study the WATER27 dataset.

\begin{figure}
    \centering
    \includegraphics[width=\linewidth]{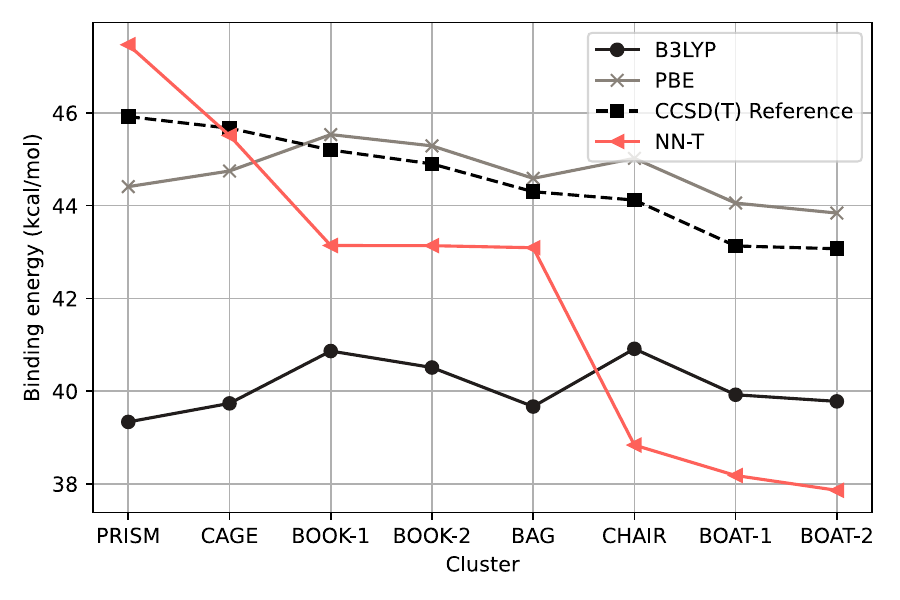}
    \caption{The binding energies of different configurations of $\mathrm{(H_2O)_6}$. Black dashed data indicates the CCSD(T) labels, while the solid lines indicate different models. Note that only the NN-T model (in orange) correctly captures the qualitative difference between different configurations.}
    \label{fig:hexamers}
\end{figure}

Finally, we compare how well different models characterize the performance of different configurations of the same system, specifically, the energy difference between the \karim{eight low-lying} hexamer configurations of water taken from Reddy \textit{et al.}~\cite{Reddy2016OnIce}. This question is of particular interest as it constitutes a key factor in the ice phase prediction while constituting a well-known and studied challenge for DFAs~\cite{Sharkas2020Self-interactionDifferences, Song2023ExtendingWater, Dasgupta2021ElevatingFormalism}. The results of our analysis are shown in figure~\ref{fig:hexamers}. We find that NN-T is nearly as accurate in mean absolute deviation on this subset as PBE, and significantly more accurate than B3LYP, PW-LDA, NN-S and NN-2, while qualitatively capturing the energy trend between different configurations, something both B3LYP and PBE fail to do correctly. \karim{Nevertheless, the absolute errors of the NN-T model remain relatively large. We conducted an investigation into 2-body and 3-body effects to assess the origins of this error. The results, presented in the supplemental material, show that the NN-T model predicts substantially weaker 2-body and 3-body interactions for these hexamers than reference calculations~\cite{Reddy2016OnIce}. This seems to indicate that the bulk of the improved energetics compared to B3LYP and LDA models arises from improved single body calculations with adjustments to two- and three-body calculations balancing the errors across different clusters.}

\section{Discussion}

Overall, we demonstrate \karim{ the ability of specialist LDA models to improve calculations on systems of interest compared to commonly used DFAs on similar rungs, with NN-S and NN-T exhibitng good performance on their respective domains}. Additionally, we demonstrate that the relevant domains are sufficiently large to enable calculations of \karim{different}  properties of interest. Crucially, we achieve these improvements with minimal data, opening the door for others to follow similar methodology and generate specialist NN-DFAs at low cost. Additional improvements in the performance of the NN-DFAs on water may be obtained by explicitly including (de-)protonated systems in the training or adding a dispersion correction. Potential future research avenues also include the usage of neural specialist DFAs for extended systems, \karim{correction for higher-order many-body effects}, and applications in \textit{ab initio} molecular dynamics.


\section{Methods}
\subsection{Calculation of molecular properties}\label{meth:mol_calc}
For each system $X$ from the ANI1-ccx~\cite{Smith2020TheMolecules} set of geometries containing atoms $X_i$ where $i$ indexes the atoms, we calculate the four single molecule properties as shown in table~\ref{tab:mol_props}. These properties were chosen as they are representative of different aspects of the electronic calculation while being trivial to calculate from a converged DFT calculation.

\begin{table}[h!]
    \centering
        \caption{Summary of molecular property calculations. $E(X)$ indicates the total energy of system $X$ from the DFT calculation.}\label{tab:mol_props}
    \begin{tabularx}{\linewidth}{|X|X|}
        \hline
        Property & Formula\\
        \hline
        Atomization E. & $E(X) - \sum_iE(X_i)$\\[2pt]
        \hline
        Ionization E. & $E(X) - E(X^+)$\\
        \hline
        Total E. & $E(X)$\\
        \hline
        LEL & See eq.~(\ref{eq:lel})\\
        \hline
    \end{tabularx}

\end{table}

\subsection{The neural density functional approximation}\label{meth:nn_dfa}

\begin{figure}[t!]
    \centering
    \includegraphics[width=.7\linewidth, clip, trim={0, 100, 0, 0}]{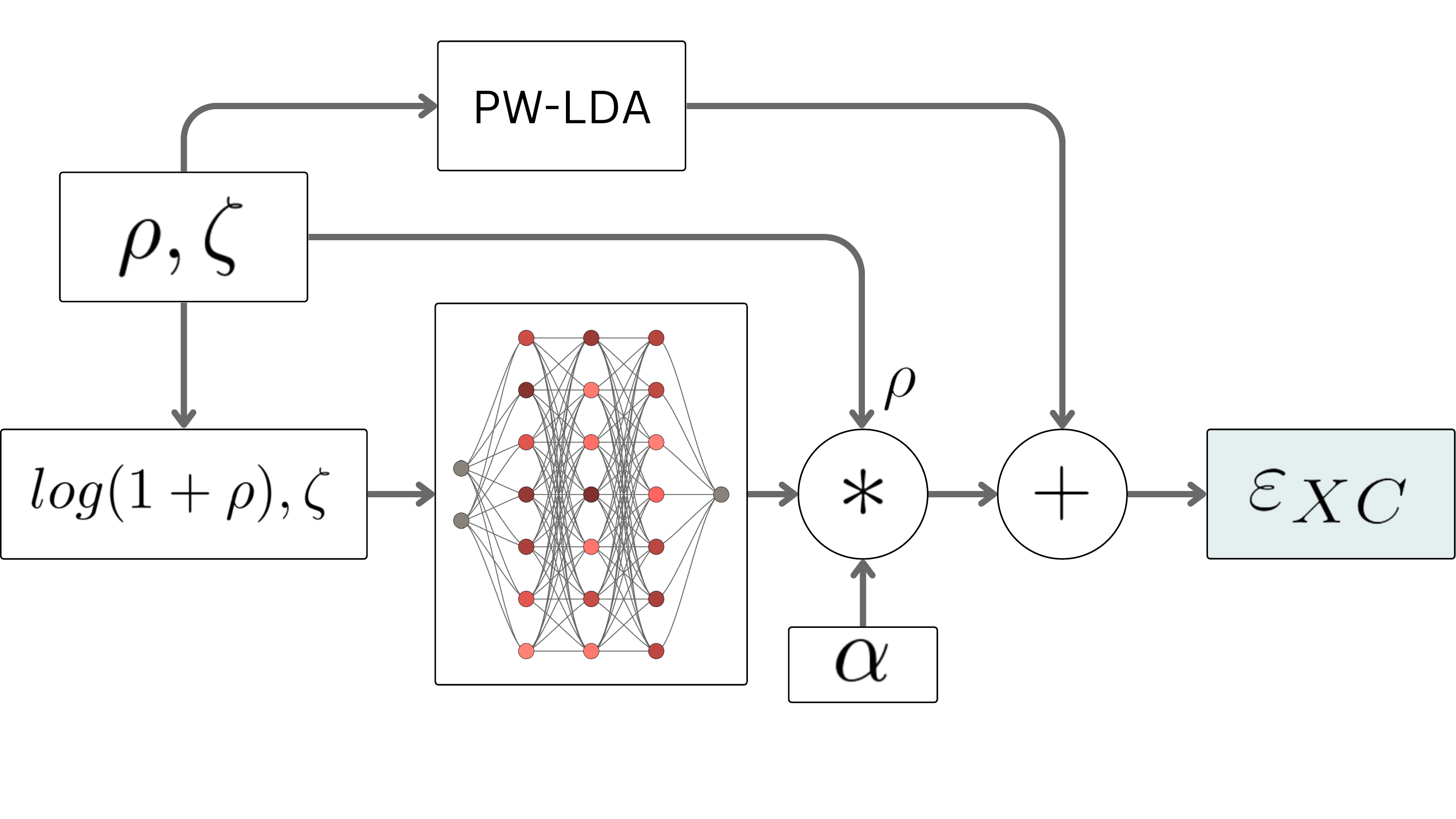}
    \caption{The delta-learning approach used to create the neural network density functional approximations (NN-DFAs). Input density $\rho$ and polarization $\zeta$ are transformed and passed to a neural network, whose output is weighted by a learnable parameter $\alpha$. This term is treated as a per-particle XC energy correction. We multiply it by $\rho$ and add it to our base model (in this case PW-LDA) and obtain an overall estimate of XC energy density $\varepsilon_{XC}$.}
    \label{fig:xc}
\end{figure}

The neural DFA presented in this work leverages delta learning starting from the spin-polarized PW-LDA, as shown in figure~\ref{fig:xc}. A neural network contribution to the local exchange-correlation energy density is calculated as 

\begin{equation}
    \varepsilon^{NN}_{XC} = \alpha\cdot \rho \cdot f(log(1 + \rho), \zeta; \theta_{NN}),
\end{equation}

where $\rho$ denotes the local electron density and $\zeta$ indicates spin polarization. The neural network function is denoted as $f$ with learnable parameters $\theta_{NN}$. In addition, $\alpha$ is a learnable parameter initialized to zero, ensuring smooth transition from the PW-LDA to the neural representation. Thus the full set of learnable parameters is

\begin{equation}
    \theta = \{\theta_{NN}, \alpha\}.
\end{equation}

The input transform $log(1 + \rho)$ was chosen to reduce the range of inputs from over five orders of magnitude to two. To maintain expressiveness of the output, we multiply the result by the density $\rho$. The parameters $\theta$ were optimized using the ADAM optimizer with a fixed learning rate of 0.005 and a mean absolute error (L1) loss.

For the neural network itself, a multi-layer perceptron (MLP) architecture was chosen. Here, the activations of a layer $\mathbf{a}_j$ are derived from the previous layer $\mathbf{a}_{j-1}$ as

\begin{equation}
    \mathbf{a}_j = \sigma\left(\mathbf{W}_j \mathbf{a}_{j-1} + \mathbf{b}_j\right).
\end{equation}

Here, $\mathbf{W}_j$ is a linear, learnable weight matrix and $\mathbf{b}_j$ is a bias vector, while $\sigma$ indicates an element-wise non-linearity.  The details of the MLP architecture are shown in table~\ref{tab:nn_params}. Increases in neural network size were not found to yield any improvements.

\begin{table}[h!]
    \centering
        \caption{Neural network hyperparameters used in training.}\label{tab:nn_params}

    \begin{tabularx}{\linewidth}{|X|X|}
        \hline
        Parameter & Value\\
        \hline
        \# Layers & 3\\
        \hline
        \# Neurons per layer & 32\\
        \hline
        $\sigma$ & Softplus\\
        \hline
    \end{tabularx}
\end{table}

\section*{Funding}
The authors acknowledge support from the Royal Society and from EPSRC under research grant EP/W010097/1.
\section*{Author contributions}
KKA and SMV conceived the project. KKA and AS developed the software and methodology, and collated the datasets used. KKA carried out simulation, training and data analysis. KKA wrote the manuscript, with contributions from all authors.
\section*{Competing interests}
The authors declare no competing interests.
\section*{Materials \& correspondence}
Materials for this study are available from the corresponding author upon reasonable request. Please contact the corresponding author with any correspondence.
\section*{Data availability}
The data that support the findings of this study can be found under the DOI \hyperlink{10.6084/m9.figshare.30657497}{https://www.doi.org/10.6084/m9.figshare.30657497}.
\section*{Code availability}
The core code this work is based on can be found under \hyperlink{https://www.github.com/OxfordHED/dqc.git}{https://www.github.com/OxfordHED/dqc.git}.
\bibliography{references}

@unpublished{Kasim2020xi-torch:Library,
    title = {{{\$}{\textbackslash}xi{\$}-torch: differentiable scientific computing library}},
    year = {2020},
    author = {Kasim, Muhammad F. and Vinko, Sam M.},
    month = {10},
    url = {https://arxiv.org/abs/2010.01921v1},
    arxivId = {2010.01921}
}

@article{Goerigk2017AInteractions,
    title = {{A look at the density functional theory zoo with the advanced GMTKN55 database for general main group thermochemistry, kinetics and noncovalent interactions}},
    year = {2017},
    journal = {Physical Chemistry Chemical Physics},
    author = {Goerigk, Lars and Hansen, Andreas and Bauer, Christoph and Ehrlich, Stephan and Najibi, Asim and Grimme, Stefan},
    number = {48},
    month = {12},
    pages = {32184--32215},
    volume = {19},
    publisher = {Royal Society of Chemistry},
    url = {https://pubs.rsc.org/en/content/articlehtml/2017/cp/c7cp04913g https://pubs.rsc.org/en/content/articlelanding/2017/cp/c7cp04913g},
    doi = {10.1039/C7CP04913G},
    issn = {14639076},
    pmid = {29110012}
}

@article{Becke1993ATheories,
    title = {{A new mixing of Hartree–Fock and local density‐functional theories}},
    year = {1993},
    journal = {The Journal of Chemical Physics},
    author = {Becke, Axel D.},
    number = {2},
    month = {1},
    pages = {1372--1377},
    volume = {98},
    publisher = {AIP Publishing},
    url = {/aip/jcp/article/98/2/1372/821432/A-new-mixing-of-Hartree-Fock-and-local-density},
    doi = {10.1063/1.464304},
    issn = {0021-9606}
}

@article{Oganov2005AbMaterials,
    title = {{Ab initio theory of planetary materials}},
    year = {2005},
    journal = {Zeitschrift fur Kristallographie},
    author = {Oganov, Artem R. and Price, G. David and Scandolo, Sandro},
    number = {5-6},
    month = {5},
    pages = {531--548},
    volume = {220},
    publisher = {De Gruyter (O)},
    url = {https://www.degruyterbrill.com/document/doi/10.1524/zkri.220.5.531.65079/html},
    doi = {10.1524/ZKRI.220.5.531.65079},
    issn = {00442968}
}

@article{Perdew1992AccurateEnergy,
    title = {{Accurate and simple analytic representation of the electron-gas correlation energy}},
    year = {1992},
    journal = {Physical Review B},
    author = {Perdew, John P. and Wang, Yue},
    number = {23},
    month = {6},
    pages = {13244},
    volume = {45},
    publisher = {American Physical Society},
    url = {https://journals.aps.org/prb/abstract/10.1103/PhysRevB.45.13244},
    doi = {10.1103/PhysRevB.45.13244},
    issn = {01631829},
    pmid = {10001404}
}

@article{Khan2025AdaptingLearning,
    title = {{Adapting hybrid density functionals with machine learning}},
    year = {2025},
    journal = {Sci. Adv},
    author = {Khan, Danish and Price, Alastair J A and Huang, Bing and Ach, Maximilian L and Anatole Von Lilienfeld, O},
    pages = {31},
    volume = {11},
    url = {https://www.science.org}
}

@article{Kulik2025AreChemist,
    title = {{Are we there yet? Adventures on a road trip through machine learning as a computational chemist}},
    year = {2025},
    journal = {APL Computational Physics},
    author = {Kulik, Heather J and Dyn, Struct},
    number = {2},
    month = {12},
    pages = {48},
    volume = {1},
    publisher = {AIP Publishing},
    url = {/aip/aco/article/1/2/020902/3366801/Are-we-there-yet-Adventures-on-a-road-trip-through},
    doi = {10.1063/5.0297853},
    issn = {3066-0017}
}

@unpublished{Akashi2025CanDFT,
    title = {{Can machines learn density functionals? Past, present, and future of ML in DFT}},
    year = {2025},
    author = {Akashi, Ryosuke and Sogal, Mihira and Burke, Kieron},
    month = {3},
    url = {https://arxiv.org/abs/2503.01709v1},
    arxivId = {2503.01709}
}

@article{Nagai2020CompletingMolecules,
    title = {{Completing density functional theory by machine learning hidden messages from molecules}},
    year = {2020},
    journal = {npj Computational Materials 2020 6:1},
    author = {Nagai, Ryo and Akashi, Ryosuke and Sugino, Osamu},
    number = {1},
    month = {5},
    pages = {1--8},
    volume = {6},
    publisher = {Nature Publishing Group},
    url = {https://www.nature.com/articles/s41524-020-0310-0},
    doi = {10.1038/s41524-020-0310-0},
    issn = {2057-3960},
    arxivId = {1903.00238}
}

@article{Schuch2009ComputationalFunctionaltheory,
    title = {{Computational complexity of interacting electrons and fundamental limitations of density functional theory}},
    year = {2009},
    journal = {Nature Physics 2009 5:10},
    author = {Schuch, Norbert and Verstraete, Frank},
    number = {10},
    month = {8},
    pages = {732--735},
    volume = {5},
    publisher = {Nature Publishing Group},
    url = {https://www.nature.com/articles/nphys1370},
    doi = {10.1038/nphys1370},
    issn = {1745-2481},
    arxivId = {0712.0483}
}

@article{vonStrachwitz2026Data-efficientDFT,
    title = {{Data-efficient learning of exchange-correlation functionals with differentiable DFT}},
    year = {2026},
    journal = {Machine Learning: Science and Technology},
    author = {von Strachwitz, Antonius and Alaa El-Din, Karim K. and Dutra, Ana C C and Vinko, Sam M},
    number = {2},
    month = {2},
    pages = {25001},
    volume = {7},
    publisher = {IOP Publishing},
    url = {https://doi.org/10.1088/2632-2153/ae3c5a},
    doi = {10.1088/2632-2153/ae3c5a}
}

@article{Fiedler2022DeepChemistry,
    title = {{Deep dive into machine learning density functional theory for materials science and chemistry}},
    year = {2022},
    journal = {Physical Review Materials},
    author = {Fiedler, L. and Shah, K. and Bussmann, M. and Cangi, A.},
    number = {4},
    month = {4},
    pages = {040301},
    volume = {6},
    publisher = {American Physical Society},
    url = {https://journals.aps.org/prmaterials/abstract/10.1103/PhysRevMaterials.6.040301},
    doi = {10.1103/PHYSREVMATERIALS.6.040301},
    issn = {24759953},
    arxivId = {2110.00997}
}

@article{Orio2009DensityTheory,
    title = {{Density functional theory}},
    year = {2009},
    journal = {Photosynthesis Research},
    author = {Orio, Maylis and Pantazis, Dimitrios A. and Neese, Frank},
    number = {2},
    month = {11},
    pages = {443--453},
    volume = {102},
    publisher = {Springer},
    url = {https://link.springer.com/article/10.1007/s11120-009-9404-8},
    doi = {10.1007/S11120-009-9404-8},
    issn = {01668595},
    pmid = {19238578}
}

@article{Vinko2014DensityPlasmas,
    title = {{Density functional theory calculations of continuum lowering in strongly coupled plasmas}},
    year = {2014},
    journal = {Nature Communications 2014 5:1},
    author = {Vinko, S. M. and Ciricosta, O. and Wark, J. S.},
    number = {1},
    month = {3},
    pages = {1--7},
    volume = {5},
    publisher = {Nature Publishing Group},
    url = {https://www.nature.com/articles/ncomms4533},
    doi = {10.1038/ncomms4533},
    issn = {2041-1723}
}

@article{Neugebauer2013DensityScience,
    title = {{Density functional theory in materials science}},
    year = {2013},
    journal = {Wiley Interdisciplinary Reviews: Computational Molecular Science},
    author = {Neugebauer, Jörg and Hickel, Tilmann},
    number = {5},
    month = {9},
    pages = {438--448},
    volume = {3},
    publisher = {John Wiley {\&} Sons, Ltd},
    url = {https://onlinelibrary.wiley.com/doi/full/10.1002/wcms.1125 https://onlinelibrary.wiley.com/doi/abs/10.1002/wcms.1125 https://wires.onlinelibrary.wiley.com/doi/10.1002/wcms.1125},
    doi = {10.1002/WCMS.1125},
    issn = {1759-0884}
}

@article{Palos2022DensityFunctional,
    title = {{Density functional theory of waterwith the machine-learned DM21 functional}},
    year = {2022},
    journal = {Journal of Chemical Physics},
    author = {Palos, Etienne and Lambros, Eleftherios and Dasgupta, Saswata and Paesani, Francesco},
    number = {16},
    month = {4},
    pages = {161103},
    volume = {156},
    publisher = {American Institute of Physics Inc.},
    url = {/aip/jcp/article/156/16/161103/2841021/Density-functional-theory-of-water-with-the},
    doi = {10.1063/5.0090862},
    issn = {10897690},
    pmid = {35490008}
}

@article{Becke1993DensityfunctionalExchange,
    title = {{Density‐functional thermochemistry. III. The role of exact exchange}},
    year = {1993},
    journal = {The Journal of Chemical Physics},
    author = {Becke, Axel D.},
    number = {7},
    month = {4},
    pages = {5648--5652},
    volume = {98},
    publisher = {AIP Publishing},
    url = {/aip/jcp/article/98/7/5648/842114/Density-functional-thermochemistry-III-The-role-of},
    doi = {10.1063/1.464913},
    issn = {0021-9606}
}

@article{Suleyman2009DFTStorage,
    title = {{DFT study of planar boron sheets: A new template for hydrogen storage}},
    year = {2009},
    journal = {Journal of Physical Chemistry C},
    author = {S{\"{u}}leyman, Er and De Wijs, Gilles A. and Brocks, Geert},
    number = {43},
    pages = {18962--18967},
    volume = {113},
    publisher = { American Chemical Society},
    url = {https://pubs.acs.org/doi/abs/10.1021/jp9077079},
    doi = {10.1021/JP9077079},
    issn = {19327447},
    arxivId = {0910.0929}
}

@article{Kasim2022DQC:Chemistry,
    title = {{DQC: A Python program package for differentiable quantum chemistry}},
    year = {2022},
    journal = {Journal of Chemical Physics},
    author = {Kasim, Muhammad F. and Lehtola, Susi and Vinko, Sam M.},
    number = {8},
    month = {2},
    pages = {84801},
    volume = {156},
    publisher = {American Institute of Physics Inc.},
    url = {/aip/jcp/article/156/8/084801/2840916/DQC-A-Python-program-package-for-differentiable},
    doi = {10.1063/5.0076202},
    issn = {10897690},
    pmid = {35232182},
    arxivId = {2110.11678}
}

@article{Dasgupta2021ElevatingFormalism,
    title = {{Elevating density functional theory to chemical accuracy for water simulations through a density-corrected many-body formalism}},
    year = {2021},
    journal = {Nature Communications 2021 12:1},
    author = {Dasgupta, Saswata and Lambros, Eleftherios and Perdew, John P. and Paesani, Francesco},
    number = {1},
    month = {11},
    pages = {6359-},
    volume = {12},
    publisher = {Nature Publishing Group},
    url = {https://www.nature.com/articles/s41467-021-26618-9},
    doi = {10.1038/s41467-021-26618-9},
    issn = {2041-1723},
    pmid = {34737311}
}

@article{Jensen2005EstimatingCalculations,
    title = {{Estimating the Hartree - Fock limit from finite basis set calculations}},
    year = {2005},
    journal = {Theoretical Chemistry Accounts},
    author = {Jensen, Frank},
    number = {5},
    month = {6},
    pages = {267--273},
    volume = {113},
    publisher = {Springer},
    url = {https://link.springer.com/article/10.1007/s00214-005-0635-2},
    doi = {10.1007/S00214-005-0635-2},
    issn = {1432881X}
}

@article{Bryantsev2009EvaluationClusters,
    title = {{Evaluation of B3LYP, X3LYP, and M06-Class density functionals for predicting the binding energies of neutral, protonated, and deprotonated water clusters}},
    year = {2009},
    journal = {Journal of Chemical Theory and Computation},
    author = {Bryantsev, Vyacheslav S. and Diallo, Mamadou S. and Van Duin, Adri C.T. and Goddard, William A.},
    number = {4},
    month = {4},
    pages = {1016--1026},
    volume = {5},
    publisher = { American Chemical Society},
    url = {https://pubs.acs.org/doi/full/10.1021/ct800549f},
    doi = {10.1021/CT800549F},
    issn = {15499618}
}

@article{Song2023ExtendingWater,
    title = {{Extending density functional theory with near chemical accuracy beyond pure water}},
    year = {2023},
    journal = {Nature Communications 2023 14:1},
    author = {Song, Suhwan and Vuckovic, Stefan and Kim, Youngsam and Yu, Hayoung and Sim, Eunji and Burke, Kieron},
    number = {1},
    month = {2},
    pages = {799-},
    volume = {14},
    publisher = {Nature Publishing Group},
    url = {https://www.nature.com/articles/s41467-023-36094-y},
    doi = {10.1038/s41467-023-36094-y},
    issn = {2041-1723},
    pmid = {36781855},
    arxivId = {2207.04169}
}

@article{Perdew1996GeneralizedSystem,
    title = {{Generalized gradient approximation for the exchange-correlation hole of a many-electron system}},
    year = {1996},
    journal = {Physical Review B},
    author = {Perdew, John P. and Burke, Kieron},
    number = {23},
    month = {12},
    pages = {16533},
    volume = {54},
    publisher = {American Physical Society},
    url = {https://journals.aps.org/prb/abstract/10.1103/PhysRevB.54.16533},
    doi = {10.1103/PhysRevB.54.16533},
    issn = {1550235X},
    pmid = {9985776}
}

@article{Perdew1996GeneralizedSimple,
    title = {{Generalized Gradient Approximation Made Simple}},
    year = {1996},
    journal = {Physical Review Letters},
    author = {Perdew, John P. and Burke, Kieron and Ernzerhof, Matthias},
    number = {18},
    month = {10},
    pages = {3865},
    volume = {77},
    publisher = {American Physical Society},
    url = {https://journals.aps.org/prl/abstract/10.1103/PhysRevLett.77.3865},
    doi = {10.1103/PhysRevLett.77.3865},
    issn = {10797114},
    pmid = {10062328}
}

@unpublished{Kuryla2025HowDatasets,
    title = {{How Accurate Are DFT Forces? Unexpectedly Large Uncertainties in Molecular Datasets}},
    year = {2025},
    author = {Kuryla, Domantas and Berger, Fabian and Cs{\'{a}}nyi, Gábor and Michaelides, Angelos and Hamied, Yusuf},
    month = {10},
    url = {https://arxiv.org/abs/2510.19774v1},
    arxivId = {2510.19774}
}

@article{Trushin2025ImprovingEnergies,
    title = {{Improving Exchange-Correlation Potentials of Standard Density Functionals with the Optimized-Effective-Potential Method for Higher Accuracy of Excitation Energies}},
    year = {2025},
    journal = {Journal of Chemical Theory and Computation},
    author = {Trushin, Egor and G{\"{o}}rling, Andreas},
    number = {4},
    month = {2},
    pages = {1667--1683},
    volume = {21},
    publisher = {American Chemical Society},
    url = {https://pubs.acs.org/doi/abs/10.1021/acs.jctc.4c01477},
    doi = {10.1021/ACS.JCTC.4C01477},
    issn = {15499626},
    pmid = {39908532}
}

@article{Sim2022ImprovingTheory,
    title = {{Improving Results by Improving Densities: Density-Corrected Density Functional Theory}},
    year = {2022},
    journal = {Journal of the American Chemical Society},
    author = {Sim, Eunji and Song, Suhwan and Vuckovic, Stefan and Burke, Kieron},
    number = {15},
    month = {4},
    pages = {6625--6639},
    volume = {144},
    publisher = {American Chemical Society},
    url = {https://pubs.acs.org/doi/abs/10.1021/jacs.1c11506},
    doi = {10.1021/JACS.1C11506},
    issn = {15205126},
    pmid = {35380807},
    arxivId = {2203.06863}
}

@article{Perdew2001JacobsEnergy,
    title = {{Jacob’s ladder of density functional approximations for the exchange-correlation energy}},
    year = {2001},
    journal = {AIP Conference Proceedings},
    author = {Perdew, John P. and Schmidt, Karla},
    number = {1},
    month = {7},
    pages = {1--20},
    volume = {577},
    publisher = {AIP Publishing},
    url = {/aip/acp/article/577/1/1/573973/Jacob-s-ladder-of-density-functional},
    doi = {10.1063/1.1390175},
    issn = {0094-243X}
}

@article{Li2021Kohn-ShamPhysics,
    title = {{Kohn-Sham Equations as Regularizer: Building Prior Knowledge into Machine-Learned Physics}},
    year = {2021},
    journal = {Physical Review Letters},
    author = {Li, Li and Hoyer, Stephan and Pederson, Ryan and Sun, Ruoxi and Cubuk, Ekin D. and Riley, Patrick and Burke, Kieron},
    number = {3},
    month = {1},
    pages = {036401},
    volume = {126},
    publisher = {American Physical Society},
    url = {https://journals.aps.org/prl/abstract/10.1103/PhysRevLett.126.036401},
    doi = {10.1103/PHYSREVLETT.126.036401},
    issn = {10797114},
    pmid = {33543980},
    arxivId = {2009.08551}
}

@article{Kasim2021LearningTheory,
    title = {{Learning the Exchange-Correlation Functional from Nature with Fully Differentiable Density Functional Theory}},
    year = {2021},
    journal = {Physical Review Letters},
    author = {Kasim, M. F. and Vinko, S. M.},
    number = {12},
    month = {9},
    volume = {127},
    publisher = {American Physical Society},
    doi = {10.1103/PHYSREVLETT.127.126403},
    issn = {10797114},
    pmid = {34597107},
    arxivId = {2102.04229}
}

@article{Dick2020MachineDensity,
    title = {{Machine learning accurate exchange and correlation functionals of the electronic density}},
    year = {2020},
    journal = {Nature Communications 2020 11:1},
    author = {Dick, Sebastian and Fernandez-Serra, Marivi},
    number = {1},
    month = {7},
    pages = {1--10},
    volume = {11},
    publisher = {Nature Publishing Group},
    url = {https://www.nature.com/articles/s41467-020-17265-7},
    doi = {10.1038/s41467-020-17265-7},
    issn = {2041-1723},
    pmid = {32665540}
}

@article{Pederson2022MachineTheory,
    title = {{Machine learning and density functional theory}},
    year = {2022},
    journal = {Nature Reviews Physics 2022 4:6},
    author = {Pederson, Ryan and Kalita, Bhupalee and Burke, Kieron},
    number = {6},
    month = {5},
    pages = {357--358},
    volume = {4},
    publisher = {Nature Publishing Group},
    url = {https://www.nature.com/articles/s42254-022-00470-2},
    isbn = {0123456789},
    doi = {10.1038/s42254-022-00470-2},
    issn = {2522-5820},
    arxivId = {2205.01591}
}

@article{Deringer2019MachineScience,
    title = {{Machine Learning Interatomic Potentials as Emerging Tools for Materials Science}},
    year = {2019},
    journal = {Advanced Materials},
    author = {Deringer, Volker L. and Caro, Miguel A. and Cs{\'{a}}nyi, Gábor},
    number = {46},
    month = {11},
    pages = {1902765},
    volume = {31},
    publisher = {John Wiley {\&} Sons, Ltd},
    url = {https://onlinelibrary.wiley.com/doi/full/10.1002/adma.201902765 https://onlinelibrary.wiley.com/doi/abs/10.1002/adma.201902765 https://advanced.onlinelibrary.wiley.com/doi/10.1002/adma.201902765},
    doi = {10.1002/ADMA.201902765},
    issn = {1521-4095},
    pmid = {31486179}
}

@article{Reddy2016OnIce,
    title = {{On the accuracy of the MB-pol many-body potential for water: Interaction energies, vibrational frequencies, and classical thermodynamic and dynamical properties from clusters to liquid water and ice}},
    year = {2016},
    journal = {Journal of Chemical Physics},
    author = {Reddy, Sandeep K. and Straight, Shelby C. and Bajaj, Pushp and Huy Pham, C. and Riera, Marc and Moberg, Daniel R. and Morales, Miguel A. and Knight, Chris and G{\"{o}}tz, Andreas W. and Paesani, Francesco},
    number = {19},
    month = {11},
    volume = {145},
    publisher = {American Institute of Physics Inc.},
    url = {/aip/jcp/article/145/19/194504/932964/On-the-accuracy-of-the-MB-pol-many-body-potential},
    doi = {10.1063/1.4967719/932964},
    issn = {00219606},
    arxivId = {1609.02884}
}

@article{Fritz2016OptimizationWater,
    title = {{Optimization of an exchange-correlation density functional for water}},
    year = {2016},
    journal = {Journal of Chemical Physics},
    author = {Fritz, Michelle and Fern{\'{a}}ndez-Serra, Marivi and Soler, José M},
    number = {22},
    month = {6},
    pages = {224101},
    volume = {144},
    issn = {0021-9606},
    arxivId = {1603.07302v2}
}

@article{Kangas1998OptimizingProperties,
    title = {{Optimizing electrostatic affinity in ligand–receptor binding: Theory, computation, and ligand properties}},
    year = {1998},
    journal = {The Journal of Chemical Physics},
    author = {Kangas, Erik and Tidor, Bruce},
    number = {17},
    month = {11},
    pages = {7522--7545},
    volume = {109},
    publisher = {AIP Publishing},
    url = {/aip/jcp/article/109/17/7522/476975/Optimizing-electrostatic-affinity-in-ligand},
    doi = {10.1063/1.477375},
    issn = {0021-9606}
}

@article{Zuo2020PerformancePotentials,
    title = {{Performance and Cost Assessment of Machine Learning Interatomic Potentials}},
    year = {2020},
    journal = {Journal of Physical Chemistry A},
    author = {Zuo, Yunxing and Chen, Chi and Li, Xiangguo and Deng, Zhi and Chen, Yiming and Behler, Jörg and Cs{\'{a}}nyi, Gábor and Shapeev, Alexander V. and Thompson, Aidan P. and Wood, Mitchell A. and Ong, Shyue Ping},
    number = {4},
    month = {1},
    pages = {731--745},
    volume = {124},
    publisher = {American Chemical Society},
    url = {https://pubs.acs.org/doi/abs/10.1021/acs.jpca.9b08723},
    doi = {10.1021/ACS.JPCA.9B08723},
    issn = {15205215},
    pmid = {31916773},
    arxivId = {1906.08888}
}

@article{Burke2012PerspectiveTheory,
    title = {{Perspective on density functional theory}},
    year = {2012},
    journal = {Journal of Chemical Physics},
    author = {Burke, Kieron},
    number = {15},
    month = {4},
    volume = {136},
    publisher = {AIP Publishing},
    url = {/aip/jcp/article/136/15/150901/941589/Perspective-on-density-functional-theory},
    doi = {10.1063/1.4704546},
    issn = {00219606},
    pmid = {22519306},
    arxivId = {1201.3679}
}

@article{Jensen2002PolarizationLimit,
    title = {{Polarization consistent basis sets. II. Estimating the Kohn–Sham basis set limit}},
    year = {2002},
    journal = {The Journal of Chemical Physics},
    author = {Jensen, Frank},
    number = {17},
    month = {5},
    pages = {7372--7379},
    volume = {116},
    publisher = {AIP Publishing},
    url = {/aip/jcp/article/116/17/7372/185286/Polarization-consistent-basis-sets-II-Estimating},
    doi = {10.1063/1.1465405},
    issn = {0021-9606}
}

@article{Jensen2001PolarizationPrinciples,
    title = {{Polarization consistent basis sets: Principles}},
    year = {2001},
    journal = {The Journal of Chemical Physics},
    author = {Jensen, Frank and Chem Phys, J},
    number = {20},
    month = {11},
    pages = {9113--9125},
    volume = {115},
    publisher = {AIP Publishing},
    url = {/aip/jcp/article/115/20/9113/442403/Polarization-consistent-basis-sets-Principles},
    doi = {10.1063/1.1413524},
    issn = {0021-9606}
}

@article{Kirkpatrick2021PushingProblem,
    title = {{Pushing the frontiers of density functionals by solving the fractional electron problem}},
    year = {2021},
    journal = {Science},
    author = {Kirkpatrick, James and McMorrow, Brendan and Turban, David H.P. and Gaunt, Alexander L. and Spencer, James S. and Matthews, Alexander G.D.G. and Obika, Annette and Thiry, Louis and Fortunato, Meire and Pfau, David and Castellanos, Lara Román and Petersen, Stig and Nelson, Alexander W.R. and Kohli, Pushmeet and Mori-S{\'{a}}nchez, Paula and Hassabis, Demis and Cohen, Aron J.},
    number = {6573},
    month = {12},
    pages = {1385--1389},
    volume = {374},
    publisher = {American Association for the Advancement of Science},
    url = {https://www.science.org/doi/10.1126/science.abj6511},
    doi = {10.1126/SCIENCE.ABJ6511},
    issn = {10959203},
    pmid = {34882476}
}

@article{Bogojeski2020QuantumLearning,
    title = {{Quantum chemical accuracy from density functional approximations via machine learning}},
    year = {2020},
    journal = {Nature Communications 2020 11:1},
    author = {Bogojeski, Mihail and Vogt-Maranto, Leslie and Tuckerman, Mark E. and M{\"{u}}ller, Klaus Robert and Burke, Kieron},
    number = {1},
    month = {10},
    pages = {1--11},
    volume = {11},
    publisher = {Nature Publishing Group},
    url = {https://www.nature.com/articles/s41467-020-19093-1},
    doi = {10.1038/s41467-020-19093-1},
    issn = {2041-1723},
    pmid = {33067479}
}

@article{Polak2025Real-spaceFunctionals,
    title = {{Real-space machine learning of correlation density functionals}},
    year = {2025},
    journal = {Nature Communications 2025 16:1},
    author = {Polak, Elias and Zhao, Heng and Vuckovic, Stefan},
    number = {1},
    month = {12},
    pages = {11306-},
    volume = {16},
    publisher = {Nature Publishing Group},
    url = {https://www.nature.com/articles/s41467-025-66450-z},
    doi = {10.1038/s41467-025-66450-z},
    issn = {2041-1723},
    pmid = {41326386}
}

@article{Lehtola2018RecentTheory,
    title = {{Recent developments in libxc — A comprehensive library of functionals for density functional theory}},
    year = {2018},
    journal = {SoftwareX},
    author = {Lehtola, Susi and Steigemann, Conrad and Oliveira, Micael J.T. and Marques, Miguel A.L.},
    month = {1},
    pages = {1--5},
    volume = {7},
    publisher = {Elsevier},
    doi = {10.1016/J.SOFTX.2017.11.002},
    issn = {2352-7110}
}

@article{Sharkas2020Self-interactionDifferences,
    title = {{Self-interaction error overbinds water clusters but cancels in structural energy differences}},
    year = {2020},
    journal = {Proceedings of the National Academy of Sciences of the United States of America},
    author = {Sharkas, Kamal and Wagle, Kamal and Santra, Biswajit and Akter, Sharmin and Zope, Rajendra R. and Baruah, Tunna and Jackson, Koblar A. and Perdew, John P. and Peralta, Juan E.},
    number = {21},
    month = {5},
    pages = {11283--11288},
    volume = {117},
    publisher = {National Academy of Sciences},
    url = {/doi/pdf/10.1073/pnas.1921258117?download=true},
    doi = {10.1073/PNAS.1921258117;WEBSITE:WEBSITE:PNAS-SITE;ISSUE:ISSUE:DOI},
    issn = {10916490},
    pmid = {32393631}
}

@article{AlaaEl-Din2024STEP:Learning,
    title = {{STEP: extraction of underlying physics with robust machine learning}},
    year = {2024},
    journal = {Royal Society Open Science},
    author = {Alaa El-Din, Karim K. and Forte, Alessandro and Kasim, Muhammad Firmansyah and Miniati, Francesco and Vinko, Sam M.},
    number = {5},
    month = {6},
    volume = {11},
    publisher = {The Royal Society},
    url = {https://royalsocietypublishing.org/doi/10.1098/rsos.231374},
    doi = {10.1098/RSOS.231374},
    issn = {20545703}
}

@article{Smith2020TheMolecules,
    title = {{The ANI-1ccx and ANI-1x data sets, coupled-cluster and density functional theory properties for molecules}},
    year = {2020},
    journal = {Scientific Data 2020 7:1},
    author = {Smith, Justin S. and Zubatyuk, Roman and Nebgen, Benjamin and Lubbers, Nicholas and Barros, Kipton and Roitberg, Adrian E. and Isayev, Olexandr and Tretiak, Sergei},
    number = {1},
    month = {5},
    pages = {1--10},
    volume = {7},
    publisher = {Nature Publishing Group},
    url = {https://www.nature.com/articles/s41597-020-0473-z},
    doi = {10.1038/s41597-020-0473-z},
    issn = {2052-4463},
    pmid = {32358545}
}

\end{document}